\documentclass{IEEEcsmag}
\usepackage{pgfplots}
\usepackage{graphics}
\usepackage{diagbox}
\usepackage{comment}
\usepackage{url}
\usepackage{makecell, array}
\usepackage{subcaption}

\tikzset{
    ultra thick/.style={line width=3.2pt}
}

\pgfplotsset{width=3cm,compat=1.8}
\usepackage{pgfplotstable}

\usepackage{sfmath}
\usepackage[binary-units]{siunitx}
\AtBeginDocument{
\DeclareSIUnit\bit{b}
}
\usepackage{multirow}
\usepackage{upmath}
\usepackage{tikz}
\usepackage[switch]{lineno}

\definecolor{dgreen}{rgb}{0.0, 0.0, 0.0}
\newcommand{\highlight}[1]{\textcolor{dgreen}{#1}}

\jvol{XX}
\jnum{XX}
\paper{8}
\jmonth{May/June}
\jname{IEEE Internet Computing}
\pubyear{2021}

\begin{document}

\title{Cloud, Fog or Edge: Where to Compute?}

\author{Dragi Kimovski, Roland Math\'{a}, Josef Hammer, Narges Mehran, Hermann Hellwagner and Radu Prodan}

\affil{Institute of Information Technology (ITEC), University of Klagenfurt}

\begin{abstract}
The computing continuum extends the high-performance cloud data centers with energy-efficient and low-latency devices close to the data sources located at the edge of the network.
However, the heterogeneity of the computing continuum raises multiple challenges related to application management. These include where to offload an application -- from the cloud to the edge -- to meet its computation and communication requirements.
To support these decisions, we provide in this article a detailed performance and carbon footprint analysis of a selection of use case applications with complementary resource requirements across the computing continuum over a real-life evaluation testbed. 
\end{abstract}
\maketitle

\textcolor{red}{\small 2021 IEEE.  Personal use of this material is permitted.  Permission from IEEE must be obtained for all other uses, in any current or future media, including reprinting/republishing this material for advertising or promotional purposes, creating new collective works, for resale or redistribution to servers or lists, or reuse of any copyrighted component of this work in other works.}

\section{1 Introduction} 
The advent of fog and edge computing has prompted predictions that they will take over the traditional cloud \highlight{for information processing and knowledge extraction} at a large scale. Notwithstanding the fact that fog and edge computing have undoubtedly large potential, these predictions are probably oversimplified and wrongly portray the relations between fog, edge and cloud computing. 
\highlight{Concretely}, fog and edge computing have been introduced as an extension of the cloud services towards the data sources, thus forming the \textit{computing continuum}.

The computing continuum enables \highlight{the creation of a new type of services, spanning} across distributed infrastructures, \highlight{for} autonomous vehicles, smart cities, and content delivery\highlight{, among other applications}. These services have a large spectrum of requirements, burdensome to meet with ``distant'' cloud data centers. For instance, they may need low-latency connections for fast decision making close to the data sources and substantial computing resources for complex data analysis. The computing continuum provides a vast heterogeneity of computational and communication resources, \highlight{which have the potential to meet these demands.} 

The heterogeneity of the computing continuum raises multiple application management challenges, such as where to offload an application from the cloud to the fog or to the edge. These issues primarily concern the large diversity of the devices, which range from single-board computers such as 
Raspberry Pis to powerful multi-processor servers. This poses the following dilemma of many practitioners and researchers:
\begin{quote}
\emph{Should we use devices accessible with low latency and with limited resource availability, or a high-performance cloud at the expense of high communication delay?}
\end{quote}

To answer this question it is essential to characterize the performance of the resources. Existing literature~\cite{sota1,sota2}, \highlight{including the DeFog benchmark suite, addresses this problem by conducting performance analysis of cloud services and to some extent of edge infrastructures. Nevertheless, these approaches ($i$) consider the edge and the cloud resources in isolation, ($ii$) provide only quantitative analysis of the performance without offloading recommendations, ($iii$) evaluate a limited number of devices, and ($iv$) do not consider the environmental impact in terms of $CO_2$ emissions for executing the applications.}

We present in this article a performance characterization and an analysis of the $CO_2$ emissions of the resources across the computing continuum. Our main goal is to \highlight{support the decision process for offloading an application to fog or edge resources by considering the application characteristics}. For this purpose, we deployed a real testbed named \emph{Carinthian Computing Continuum} ($C^3$) that aggregates a large set of heterogeneous resources. We base the analysis on three complementary applications widely utilized by industry and research: video encoding, machine learning and in-memory data analytics. We conclude by providing recommendations on where to compute applications across the computing continuum.

\section{2 Carinthian Computing Continuum}
\begin{figure}[t]
	\centering
    \includegraphics[width=\columnwidth]{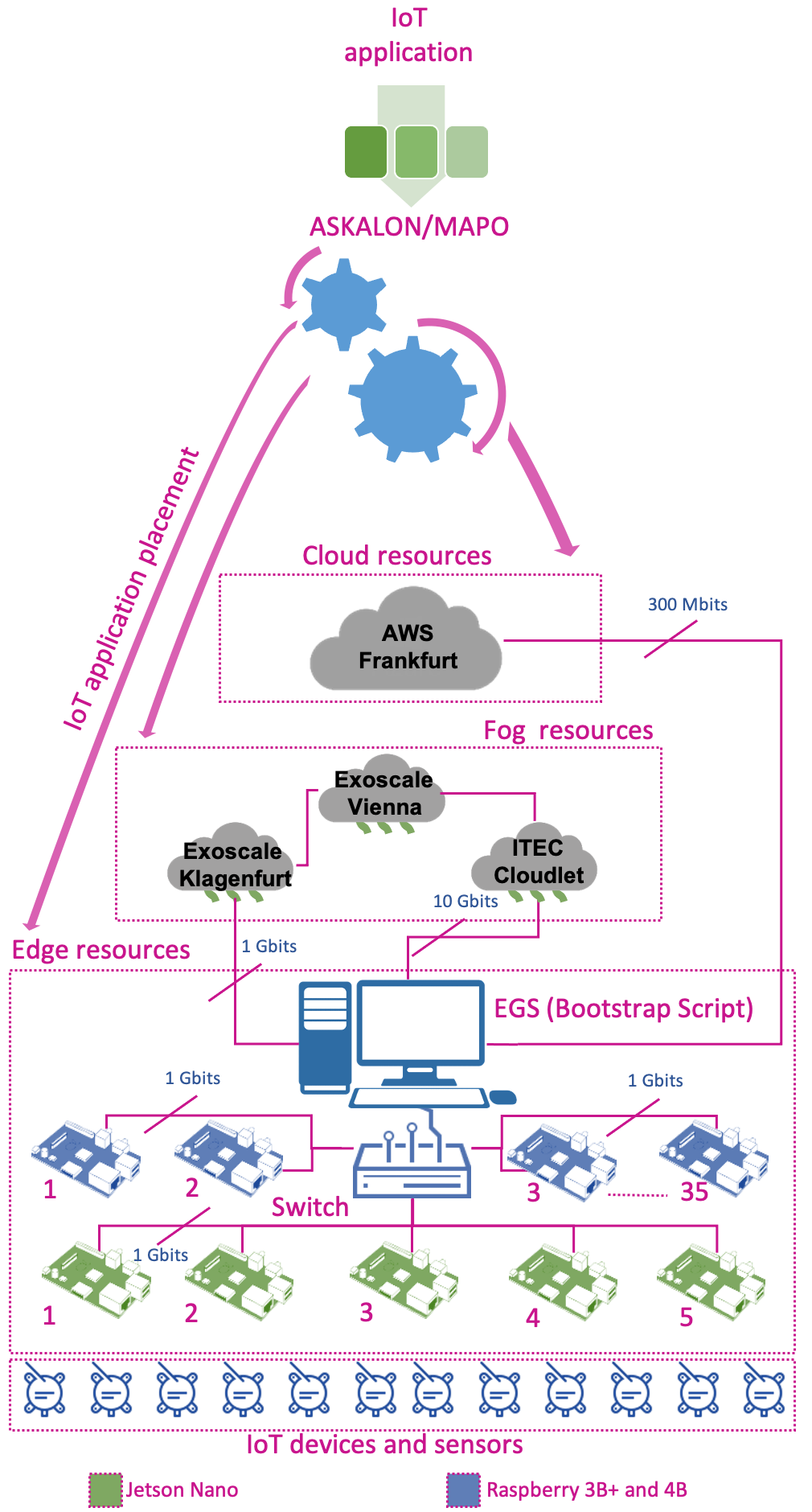}
	\caption{The $C^3$ testbed.}
    \label{figArch}
\end{figure}

Figure \ref{figArch} depicts the top-level view of the Carinthian Computing Continuum. The $C^3$ testbed includes a heterogeneous set of resources, distributed across different control domains, including public providers such as Exoscale Cloud\footnote{\url{https://www.exoscale.com}} and Amazon Web Services (AWS), and research institutions such as University of Klagenfurt\footnote{\url{https://itec.aau.at}}. \highlight{We utilize the ASKALON cloud application computing environment~\cite{a3} with the MAPO resource provisioning algorithm~\cite{a2} to deploy the applications across the $C^3$ testbed. Furthermore, we employ a bootstrapping script that automatically configures the resources in the testbed \footnote{\url{https://github.com/josefhammer/c3-edge}}.  
Table \ref{tbl:InstanceTypes} summarizes the resource characteristics of the $C^3$ testbed.}  

\highlight{We classify the resources in the $C^3$ testbed into three layers: \emph{cloud layer}, \emph{fog layer} and \emph{edge layer}.} 

\begin{table*}[t]
\scriptsize
\caption{Description of the resources available in the $C^3$ testbed.}
\centering
{
\resizebox{\textwidth}{!}{
\begin{tabular}{|@{ }c@{ }|@{ }c@{ }|@{ }c@{ }|@{ }r@{ }|@{ }r@{ }|@{ }c@{ }|@{ }c@{ }|@{ }c@{ }|@{ }c@{ }|@{ }c@{ }|}
\hline
 \textit{Conceptual layer} &
 \textit{Device / Instance type} & \textit{Architecture}& \textit{(v)CPU} & \textit{Memory [\SI{}{\gibi\byte}]} & \textit{Storage [\SI{}{\gibi\byte}]} & \textit{Network} & \textit{Physical processor} & \textit{Clock [\SI{}{\giga\hertz}]}& \textit{Operating system}\\
\hline
\multirow{3}{*}{Cloud layer} &\texttt{AWS t2.micro}& \multirow{3}{*}{64-bit x86} & 1 & 1 & \multirow{3}{*}{32} &  Moderate & Intel Xeon  & $\le 3.1$&\multirow{3}{*}{Ubuntu 18.04}\\
& \texttt{AWS c5.large}& &2 & 4 & &\multirow{2}{*}{$\le 10$ \SI{}{\giga\bit\per\second}} & Intel Xeon Platinum 8000 series& $\le 3.6$&\\
& \texttt{AWS m5a.xlarge}& &4 & 16 & & & AMD EPYC 7000 series & $\le 2.5$&\\
\hline
\multirow{4}{*}{Fog layer} & \texttt{Exoscale Tiny}& \multirow{4}{*}{64-bit x86} & 1 & 1 & \multirow{4}{*}{32} &  \multirow{4}{*}{$\le 10$ \SI{}{\giga\bit\per\second}} &
\multirow{3}{*}{Intel Xeon}  & 
\multirow{3}{*}{$\le 3.6$}&
\multirow{4}{*}{Ubuntu 18.04}\\
& \texttt{Exoscale Medium}& &2 & 4 & & & & &\\
&\texttt{Exoscale Large}& &4 & 8 & & & & &\\
& \texttt{ITEC Cloud Instance}&  & 4 & 8 &  &  & Intel Xeon Platinum 8000  & $\le 3.1$&\\
\hline
\multirow{4}{*}{Edge layer} & \texttt{Edge Gateway System}& 64-bit x86 & 12 & 32 & 32 &  $\le 10$ \SI{}{\giga\bit\per\second} & AMD Ryzen Threadripper 2920X  & $\le 3.5$ & Ubuntu 18.04\\
&\texttt{Raspberry Pi 3B}& \multirow{3}{*}{64-bit ARM} &\multirow{3}{*}{4} & 1 &  \multirow{3}{*}{64} &\multirow{3}{*}{$\le 1$ \SI{}{\giga\bit\per\second}} & Cortex - A53 & $\le 1.4$ & \multirow{2}{*}{Pi OS Buster}\\
&\texttt{Raspberry Pi 4}& & & 4 & & & Cortex - A72 & $\le 1.5$&\\
&\texttt{Jetson Nano}& & & 4 & & & Tegra X1 and Cortex - A57 & $\le 1.43$& Linux for Tegra R28.2.1\\
\hline
\end{tabular}
}
}
\label{tbl:InstanceTypes}
\end{table*}

\subsection{2.1 Cloud layer} The cloud layer is the uppermost layer of the $C^3$ testbed. It contains high-performance resources consolidated in vast data-centers, provisioned on-demand as virtual machine instances. As the $C^3$ testbed resides in Klagenfurt (Austria), we complement it with the geographically closest European AWS cloud data center located in Frankfurt (Germany).

We carefully selected three instance types based on the \texttt{x86-64} architecture that offer to the $C^3$ testbed a balance of compute, memory, and networking resources for a broad set of applications:
general purpose (\texttt{t2.micro}), and compute-optimized (\texttt{c5.large} and \texttt{m5a.xlarge}).

\subsection{2.2 Fog layer} \highlight{The fog layer comprises computing infrastructures consolidated in small data-centers in close vicinity to the data sources.} This layer comprises resources from two providers in the $C^3$ testbed~\cite{a1}: Exoscale and University of Klagenfurt. We allocate these providers in the fog  layer as a result of the low round-trip communication latency ($\leq$ \SI{7}{\milli\second}) and high bandwidth ($\leq$ \SI{10}{\giga\bit\per\second}). The Exoscale cloud comprises data centers in Vienna and Klagenfurt (Austria). We selected three computing optimized \texttt{x86-64} instances from the Exoscale cloud offering: \texttt{Tiny}, \texttt{Medium} and \texttt{Large}. 
University of Klagenfurt provides a private cloud infrastructure operated by \texttt{OpenStack v13.0} and \texttt{Ceph v12.2}
with one computing optimized instance type described in Table \ref{tbl:InstanceTypes}.

\subsection{2.3 Edge layer} 
This layer encompasses edge resources, such as single-board computers, directly connected to the IoT devices and sensors. An Edge Gateway System (EGS) controls the edge layer, and is the entry point to the other resources available on this level. The EGS supports  \SI{10}{\giga\bit\per\second} Ethernet, dual band PCIe WiFi 5 (802.11ac) and a \SI{150}{\mega\bit\per\second} LTE \SI{2600}{\mega\hertz} connection. A layer-3 HP Aruba switch with 48 \SI{1}{\giga\bit\per\second} ports connects the EGS to the single-board computers with a latency of \SI{3.8}{\micro\second} and an aggregate data transfer rate of \SI{104}{\giga\bit\per\second}. The edge layer also contains 35 physical nodes based on either Raspberry Pi 3B or Pi 4B.  Besides, the testbed contains five Jetson Nano devices, each equipped with a general purpose GPU.
The edge layer has \SI{1}{\giga\bit\per\second} Ethernet, Wi-Fi and LTE network connection interfaces.

\section{3 Benchmark applications}
We selected three representative application classes with complementary requirements to evaluate the computational performance and the $CO_2$ emissions of the computing continuum. 

\subsection{3.1 Video encoding} Video encoding allows transmission of video content with different qualities over limited and heterogeneous communication channels. It compresses an original raw video to reduce its effective bandwidth consumption, while maintaining a subjective high quality for viewers. Video encoding has wide fields of applications, including content delivery (live and on-demand video streams), traffic control and surveillance. \highlight{The video encoding applications have high processing and throughput requirements.} 

\subsection{3.2 Machine learning} Machine learning is a branch of artificial intelligence that explores approaches for enabling systems to learn from data, identify patterns and make decisions. Its vast field of application includes automated control in manufacturing, adaptive traffic planning and smart health-care diagnosis, among others. \highlight{Machine learning, in general, has high processing and operating memory requirements.} 

\subsection{3.3 In-memory analytic} In-memory analytic is essential for efficient low-latency decisions on devices with limited resources. It explores data manipulation such as inspecting, ﬁltering and transforming, and enables efficient extraction of knowledge and non-biased decision-making. Its fields of application include smart cities, healthcare and recommender systems. \highlight{The in-memory analytic applications require large memory capacity and strict communication latency.} 

\section{\highlight{4 Performance evaluation}}
\subsection{4.1 Video encoding}
We evaluate the encoding performance of the computing continuum using FFmpeg version 3.4.6 with the most popular H.264/MPEG-4 video encoder\footnote{\url{https://trac.ffmpeg.org/wiki/Encode/H.264}} deployed by more than $90\%$ of the video industry\footnote{\url{https://www.itu.int/rec/T-REC-H.264-201906-I/en}}. We perform the encoding on a raw video segment with length of \SI{4}{\second} and size of  \SI{514}{\mega\byte}, available in the Sintel\footnote{\url{https://media.xiph.org/sintel}} video-set. The video segment is encoded in three resolutions (HD-ready, Full HD and Quad HD) with data rates of 1500, 3000, and 6500 $\SI{}{\kilo\bit\per\second}$. 

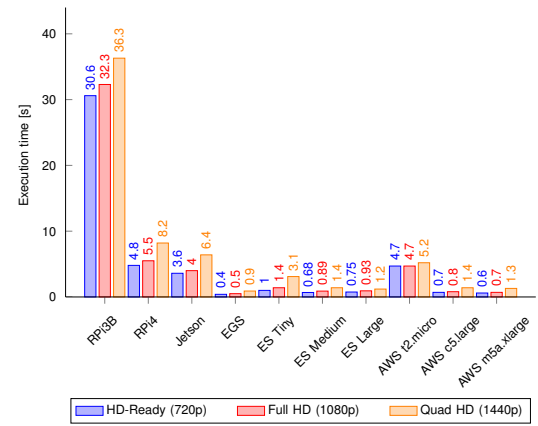
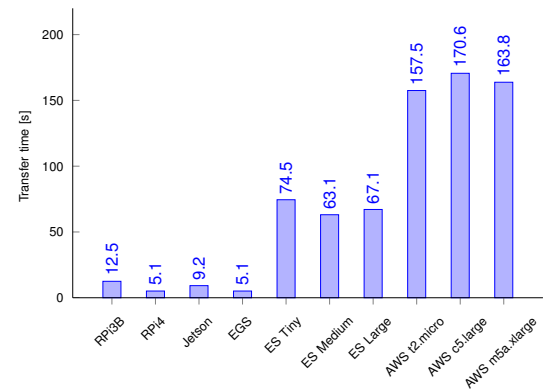
\begin{figure}[h!]
    \centering
\begin{subfigure}[t]{\columnwidth}    
\begin{tikzpicture}[scale=0.6]
  \centering
  \begin{axis}[
        ybar, axis on top,
        height=8cm, width=12cm,
        bar width=0.25cm,
        nodes near coords,
        every node near coord/.append style={rotate=90, anchor=west, font=\footnotesize},
       major grid style={draw=black},
        enlarge y limits={value=.1,upper},
        ymin=0, ymax=40,
        legend style={
            at={(0.5,-0.35)},
            anchor=north,
            legend columns=-1,
            font=\footnotesize,
            /tikz/every even column/.append style={column sep=0.6cm}
        },
        restrict y to domain*=0:117, 
    visualization depends on=rawy\as\rawy, 
    after end axis/.code={ 
            \draw [ultra thick, white, decoration={snake, amplitude=2pt}, decorate] (rel axis cs:0,1.01) -- (rel axis cs:1,1.01);
        },
        nodes near coords={%
            \pgfmathprintnumber{\rawy}
        },
        ylabel={Execution time [\SI{}{\second}]},
        y label style = {xshift=-0.2em,font=\footnotesize},
         yticklabel style = {xshift=-0.2em,font=\footnotesize},
        symbolic x coords={
            RPi3B, RPi4, Jetson, EGS, ES Tiny, ES Medium, ES Large, AWS t2.micro, AWS c5.large, AWS m5a.xlarge},
       xticklabel style = {rotate=45, xshift=-0.2em,font=\footnotesize},
       axis lines*=left,
         clip=false,
         area legend
    ]
    \addplot [color =blue, fill=blue!30] coordinates {
      
      (RPi3B, 30.6)
       (RPi4, 4.8)
        (Jetson, 3.6)
      (EGS, 0.4)
      (ES Tiny, 1.0) 
      (ES Medium, 0.68)
      (ES Large, 0.75) 
      (AWS t2.micro, 4.7)
      (AWS c5.large, 0.7) 
       (AWS m5a.xlarge, 0.6)
       };

        \addplot [color=red, fill=red!30] coordinates {
       
      (RPi3B, 32.3) 
       (RPi4, 5.5)
       (Jetson, 4.0)
      (EGS, 0.5)
      (ES Tiny, 1.4) 
      (ES Medium, 0.89)
      (ES Large, 0.93) 
      (AWS t2.micro, 4.7)
      (AWS c5.large, 0.8) 
       (AWS m5a.xlarge, 0.7)
       };

               \addplot [color=orange, fill=orange!30] coordinates {
       
      (RPi3B, 36.3) 
       (RPi4, 8.2)
       (Jetson, 6.4)
      (EGS, 0.9)
      (ES Tiny, 3.1) 
      (ES Medium, 1.4)
      (ES Large, 1.2) 
      (AWS t2.micro, 5.2)
      (AWS c5.large, 1.4) 
       (AWS m5a.xlarge, 1.3)
       };
    \legend{HD-Ready (720p), Full HD (1080p), Quad HD (1440p)}
  \end{axis}
  
  \end{tikzpicture}
    \caption{Average encoding time.}
    \label{fig:encode}
\end{subfigure}   
\begin{subfigure}[t]{\columnwidth}    
\begin{tikzpicture}[scale=0.6]
  \centering
  \begin{axis}[
        ybar, axis on top,
        height=8cm, width=12cm,
        bar width=0.4cm,
        nodes near coords,
        every node near coord/.append style={rotate=90, anchor=west},
       major grid style={draw=black},
        enlarge y limits={value=.1,upper},
        ymin=0, ymax=200,
        legend style={
            at={(0.5,-0.3)},
            anchor=north,
            legend columns=-1,
            font=\footnotesize,
            /tikz/every even column/.append style={column sep=0.5cm}
        },
        restrict y to domain*=0:250, 
    visualization depends on=rawy\as\rawy, 
    after end axis/.code={ 
           \draw [ultra thick, white,  decoration={snake, amplitude=2pt}, decorate] (rel axis cs:0,1.01) -- (rel axis cs:1,1.01);
        },
        nodes near coords={%
            \pgfmathprintnumber{\rawy}
        },
        ylabel={Transfer time [\SI{}{\second}]},
        y label style = {xshift=-0.2em,font=\footnotesize},
         yticklabel style = {xshift=-0.2em,font=\footnotesize},
        symbolic x coords={
            RPi3B, RPi4, Jetson, EGS, ES Tiny, ES Medium, ES Large, AWS t2.micro, AWS c5.large, AWS m5a.xlarge},
       xticklabel style = {rotate=45, xshift=-0.2em,font=\footnotesize},
       axis lines*=left,
         clip=false,
         area legend
    ]
    \addplot [color =blue, fill=blue!30] coordinates {
     
      (RPi3B, 12.5)
       (RPi4, 5.1)
       (Jetson, 9.2)
      (EGS, 5.1)
      (ES Tiny, 74.5) 
      (ES Medium, 63.1)
      (ES Large, 67.1) 
      (AWS t2.micro, 157.5)
      (AWS c5.large,170.6) 
       (AWS m5a.xlarge, 163.8)
       };
  \end{axis}
  
  \end{tikzpicture}
    \caption{\highlight{Average raw video segment transfer time.}}
    \label{fig:trans}
\end{subfigure}  
    \caption{Average encoding performance of a \SI{4}{\second} long video segment with the x264 codec and FFmpeg 3.4.6.}
    \label{fig:encodeWhole}
\end{figure}

Figure \ref{fig:encodeWhole} depicts the average encoding time and \highlight{transfer time, from the video source (located at the University of Klagenfurt) to the encoding device or instance,} for a single raw video segment in the three resolutions.  The standard deviation ranges from $1.3\%$ for the AWS \texttt{m5a.xlarge} instance to $3.6\%$ for the Raspberry Pi 3B devices. We observe that the older generation single-board computers (Raspberry Pi 3B) have a significantly higher encoding time than the other resources. 
\highlight{However, the Raspberry Pi 3B devices provide lower transfer times than the cloud instances and are suitable for video-on-demand services employing offline encoding.} \highlight{The Raspberry Pi 4 and the Jetson Nano devices efficiently perform video encoding and provide low transfer times.} In some cases, Jetson Nano was capable of encoding up to 20\% faster than the AWS \texttt{t2.micro} instance \highlight{with significantly lower transfer times.} The remaining cloud and fog resources showed similarly video encoding performance in the range between \SIrange{0.5}{1.3}{\second}. \highlight{Nevertheless, the cloud and fog resources have limited effective throughput causing higher raw video transfer times. However, the cloud resources are suitable for live video streaming due to the low encoding times. Overall,} the EGS achieved the lowest encoding and transfer time due to the low utilization rate and its high computing and networking capabilities.

\subsubsection{Recommendation.} We recommend executing video-on-demand encoding at the edge using the latest generation of single-board computers or dedicated systems, \highlight{as they significantly reduce the raw video transfer time}. Cloud and fog devices (i.e., close-by servers, small data centers) are more suitable for \highlight{continuous live stream encoding} if the effective incoming and outgoing throughput is sufficient and the delay incurred by the transport is tolerable.

\subsection{4.2 Machine learning}
We use TensorFlow Core version 2.3.0 to evaluate machine learning performance. We created two training and validation scenarios for feature identification in a set of images:
\begin{itemize}
\item A \emph{quantum neural network} using the MNIST data-set\footnote{\url{http://yann.lecun.com/exdb/mnist/}} limited to $20000$ samples with a size of \SI{3.3}{\mega\byte}. The scenario creates a neural network with two layers and 128 outputs from the previous layer to the next. We conduct five iterations to reach a feature identification accuracy of $90\%$.
\item A \emph{convolutional neural network} using the Kaggle data-set\footnote{\url{https://www.kaggle.com/tags/animals}} with a size of \SI{218}{\mega\byte}. The minimum required accuracy is $80\%$. The convolutional network has three layers with a kernel size of three. Each layer uses increasingly higher filter sizes in the range [32, 64, 128]. After each layer, we use a max-pooling sample-based discretization process to reduce the spatial dimensions. We repeat the training five times.
\end{itemize}

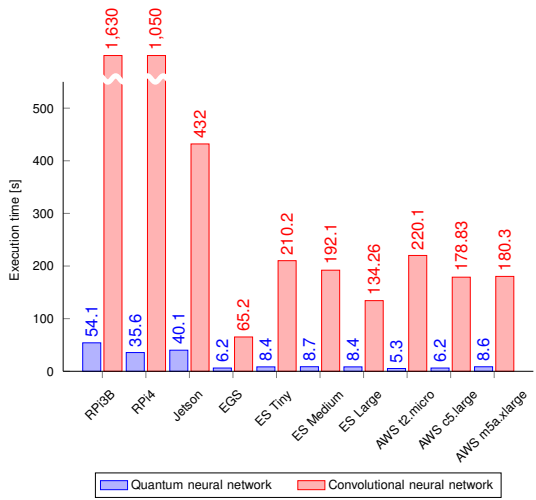
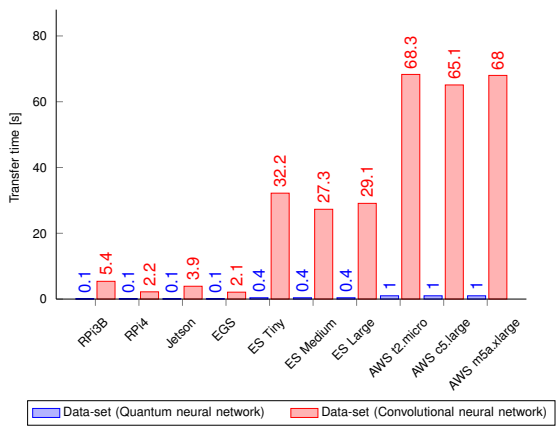
\begin{figure}[h!]
    \centering
\begin{subfigure}[t]{\columnwidth}     
\begin{tikzpicture}[scale=0.6]
  \centering
  \begin{axis}[
        ybar, axis on top,
        height=8cm, width=12cm,
        bar width=0.4cm,
        nodes near coords,
        every node near coord/.append style={rotate=90, anchor=west},
       major grid style={draw=black},
        enlarge y limits={value=.1,upper},
        ymin=0, ymax=500,
       restrict y to domain*=0:600, 
    visualization depends on=rawy\as\rawy, 
    after end axis/.code={ 
            \draw [ultra thick, white, decoration={snake, amplitude=2pt}, decorate] (rel axis cs:0,1.01) -- (rel axis cs:1,1.01);
        },
        nodes near coords={%
            \pgfmathprintnumber{\rawy}
        },
        legend style={
            at={(0.5,-0.35)},
            anchor=north,
            legend columns=-1,
            font=\footnotesize,
            /tikz/every even column/.append style={column sep=0.5cm}
        },
        ylabel={Execution time [\SI{}{\second}]},
        y label style = {xshift=-0.2em,font=\footnotesize},
         yticklabel style = {xshift=-0.2em,font=\footnotesize},
        symbolic x coords={
            RPi3B,RPi4, Jetson, EGS, ES Tiny, ES Medium, ES Large, AWS t2.micro, AWS c5.large, AWS m5a.xlarge},
       xticklabel style = {rotate=45, xshift=-0.2em,font=\footnotesize},
         axis lines*=left,
    clip=false,
    area legend
    ]
    \addplot [color =blue, fill=blue!30] coordinates {
      (RPi3B, 54.1) 
      (RPi4, 35.6)
      (Jetson, 40.1) 
      (EGS, 6.2)
      (ES Tiny, 8.4) 
      (ES Medium, 8.7)
      (ES Large, 8.4) 
      (AWS t2.micro, 5.3)
      (AWS c5.large, 6.2) 
       (AWS m5a.xlarge, 8.6)
       };
       
        \addplot [color=red, fill=red!30] coordinates {
      (RPi3B, 1630)
      (RPi4, 1050)
      (Jetson, 432)
      (EGS, 65.2)
      (ES Tiny, 210.2) 
      (ES Medium, 192.1)
      (ES Large, 134.26) 
      (AWS t2.micro, 220.1)
      (AWS c5.large,178.83) 
       (AWS m5a.xlarge,180.3)
       };  
    \legend{Quantum neural network, Convolutional neural network}
  \end{axis}
  
  \end{tikzpicture}
      \caption{Average training time.}
    \label{fig:trans}
  \end{subfigure}   
\begin{subfigure}[t]{\columnwidth}    
\begin{tikzpicture}[scale=0.6]
  \centering
  \begin{axis}[
        ybar, axis on top,
        height=8cm, width=12cm,
        bar width=0.4cm,
        nodes near coords,
        every node near coord/.append style={rotate=90, anchor=west},
       major grid style={draw=black},
        enlarge y limits={value=.1,upper},
        ymin=0, ymax=80,
        legend style={
            at={(0.5,-0.35)},
            anchor=north,
            legend columns=-1,
            font=\footnotesize,
            /tikz/every even column/.append style={column sep=0.5cm}
        },
        restrict y to domain*=0:110, 
    visualization depends on=rawy\as\rawy, 
    after end axis/.code={ 
           \draw [ultra thick, white,  decoration={snake, amplitude=2pt}, decorate] (rel axis cs:0,1.01) -- (rel axis cs:1,1.01);
        },
        nodes near coords={%
            \pgfmathprintnumber{\rawy}
        },
        ylabel={Transfer time [\SI{}{\second}]},
        y label style = {xshift=-0.2em,font=\footnotesize},
         yticklabel style = {xshift=-0.2em,font=\footnotesize},
        symbolic x coords={
            RPi3B, RPi4, Jetson, EGS, ES Tiny, ES Medium, ES Large, AWS t2.micro, AWS c5.large, AWS m5a.xlarge},
       xticklabel style = {rotate=45, xshift=-0.2em,font=\footnotesize},
       axis lines*=left,
         clip=false,
         area legend
    ]
    \addplot [color =blue, fill=blue!30] coordinates {
     
      (RPi3B, 0.1)
       (RPi4, 0.1)
       (Jetson, 0.1)
      (EGS, 0.1)
      (ES Tiny, 0.4) 
      (ES Medium, 0.4)
      (ES Large, 0.4) 
      (AWS t2.micro, 1.0)
      (AWS c5.large,1.0) 
       (AWS m5a.xlarge, 1.0)
       };
          \addplot [color=red, fill=red!30] coordinates {
     
      (RPi3B, 5.4)
       (RPi4, 2.2)
       (Jetson, 3.9)
      (EGS, 2.1)
      (ES Tiny, 32.2) 
      (ES Medium, 27.3)
      (ES Large, 29.1) 
      (AWS t2.micro, 68.3)
      (AWS c5.large, 65.1) 
       (AWS m5a.xlarge, 68.0)
       };
         \legend{Data-set (Quantum neural network), Data-set (Convolutional neural network)}
  \end{axis}
  
  \end{tikzpicture}
    \caption{\highlight{Average training data transfer time.}}
    \label{fig:trans}
\end{subfigure}

    \caption{Average training and data transfer times of two neural network types.}
    \label{fig:tensor}
\end{figure}

Figure~\ref{fig:tensor} analyzes the average execution time for training the two neural network types \highlight{and the transfer times of the training data from centralized storage to the device or instance that performs the training.} The standard deviation ranges from $1.2\%$ for the Raspberry Pi 4 devices to $5.4\%$ for the AWS \texttt{t2.micro} instance. The evaluation shows that the less complex quantum neural network requires a relatively lower training time across all resources. The old generation single-board computers show again a lower performance, and their suitability for training heavily depends on the size of the training data and the model. The other fog and edge devices provide similar performance to the cloud resources. The  single-board computers provide lower training performance for the convolutional network. The only exception are the Jetson Nano devices able to train the convolutional network up to four times faster than the Raspberry Pi devices. In general, the EGS provides the lowest training time among all devices. \highlight{The training data transfer time has limited influence on the training process, especially for the quantum neural network. While the training data transfer time is significantly higher for the convolutional neural network, the cloud and fog resources outperform the edge devices, except EGS.} 

\subsubsection{Recommendation.} We recommend the model training with large data-sets and multiple layers in the cloud or on dedicated systems (such as EGS), whenever possible. We recommend offloading to the edge only when the training data is of limited size, or when the neural network has few layers.

\subsection{4.3 In-memory data analytics}
The in-memory data analytics evaluation explores two scenarios using Apache Spark version 2.4.6:
\begin{itemize}
\item \emph{Collaborative data filtering} aims to fill missing entries for improved recommendation of movies to consumers. The model uses the alternating least squares algorithm and a data-set\footnote{\url{https://github.com/apache/spark/blob/master/data/mllib/als/sample_movielens_ratings.txt}} of movie preferences with a size of \SI{31.6}{\kilo\byte}. We trained the model over the available data-set with a cold start strategy that randomly divides the data into training and validation sets.
\item \emph{$\pi$ estimation} is a memory and computationally intensive task that estimates the value of $\pi$ by distributing the work among multiple Spark executors. This enables us to evaluate the computational and memory performance of the distributed memory computing continuum for complex tasks.
\end{itemize}

\begin{figure}[h!]
    \centering
\begin{tikzpicture}[scale=0.6]
  \centering
  \begin{axis}[
        ybar, axis on top,
        height=8cm, width=12cm,
        bar width=0.4cm,
        nodes near coords,
        every node near coord/.append style={rotate=90, anchor=west},
       major grid style={draw=black},
        enlarge y limits={value=.1,upper},
        ymin=0, ymax=100,
        legend style={
            at={(0.5,-0.35)},
            anchor=north,
            legend columns=-1,
            font=\footnotesize,
            /tikz/every even column/.append style={column sep=0.5cm}
        },
        restrict y to domain*=0:117, 
    visualization depends on=rawy\as\rawy, 
    after end axis/.code={ 
            \draw [ultra thick, white, decoration={snake, amplitude=2pt}, decorate] (rel axis cs:0,1.01) -- (rel axis cs:1,1.01);
        },
        nodes near coords={%
            \pgfmathprintnumber{\rawy}
        },
        ylabel={Execution time [\SI{}{\second}]},
        y label style = {xshift=-0.2em,font=\footnotesize},
         yticklabel style = {xshift=-0.2em,font=\footnotesize},
        symbolic x coords={
            RPi3B, RPi4, Jetson, EGS, ES Tiny, ES Medium, ES Large, AWS t2.micro, AWS c5.large, AWS m5a.xlarge},
       xticklabel style = {rotate=45, xshift=-0.2em,font=\footnotesize},
       axis lines*=left,
         clip=false,
         area legend
    ]
    \addplot [color =blue, fill=blue!30] coordinates {

           (RPi3B, 144.1) 
      (RPi4, 77.3)
      (Jetson, 81.1)
      (EGS, 46)
      (ES Tiny, 38.8) 
      (ES Medium, 20.6)
      (ES Large, 15.05) 
      (AWS t2.micro, 52.7)
      (AWS c5.large, 25.1) 
       (AWS m5a.xlarge, 29.2)
       };  
       
        \addplot [color=red, fill=red!30] coordinates {
           (RPi3B, 16.5)
      (RPi4, 7.4)
      (Jetson, 6.8)
      (EGS, 4.73)
      (ES Tiny, 2.19) 
      (ES Medium, 1.69)
      (ES Large, 1.57) 
      (AWS t2.micro, 3.9)
      (AWS c5.large, 2.1) 
       (AWS m5a.xlarge, 2.5)
       };  
  
    \legend{Collaborative filtering , Pi Calculation}
  \end{axis}
  
  \end{tikzpicture}
    \caption{Average execution time for in-memory collaborative data filtering and $\pi$ estimation using Apache Spark.}
    \label{fig:spark}
\end{figure}
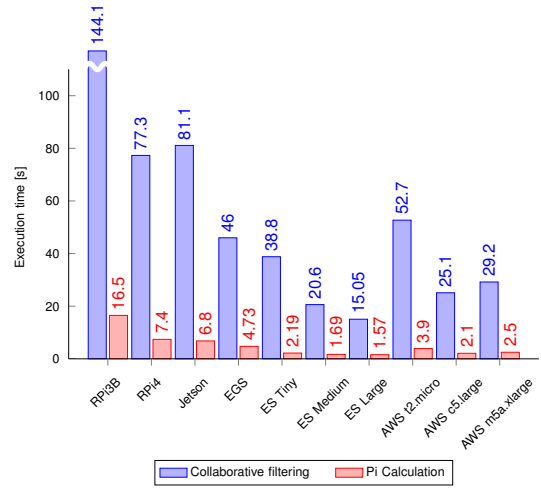

Figure~\ref{fig:spark} shows the average execution time of the in-memory collaborative data filtering and the $\pi$ estimation. The standard deviation ranges from $1.3\%$ for the AWS \texttt{m5a.xlarges} instance to $4.6\%$ for the Exoscale \texttt{Tiny} instance.
The AWS and Exoscale cloud instances perform better than the EGS and the single-board computers for the $\pi$ calculation thanks to their larger memory size and the more efficient memory controllers. The collaborative filtering shows the same trend and the Exoscale instances in Vienna show the best performance. \highlight{The data transfer time of the collaborative filtering is negligible due to its small size.}

\subsubsection{Recommendation.} \highlight{We recommend fog instances for collaborative data filtering, due to the relatively small difference in the execution time compared to the cloud. The edge devices can be a reasonable option for applications with soft constraints on the data filtering time.} Finally, we recommend executing compute intensive in-memory processing (e.g., $\pi$ estimation) in the cloud or offloading to fog devices with good memory management.

\subsection{4.4 Network performance}
\highlight{Furthermore, we evaluate the network performance} of each instance and device in the $C^3$ testbed by measuring the effective downlink throughput with the iPerf3\footnote{\url{https://iperf.fr/}} tool over TCP and the round-trip latency by sending ICMP echo requests from a device registered in the University of Klagenfurt network.

\begin{figure}[h!]
    \centering
\begin{tikzpicture}[scale=0.6]
  \centering
  \begin{axis}[
        ybar, axis on top,
        height=8cm, width=12cm,
        bar width=0.4cm,
        nodes near coords,
        every node near coord/.append style={rotate=90, anchor=west},
       major grid style={draw=black},
        enlarge y limits={value=.1,upper},
        ymin=0, ymax=190,
        legend style={
            at={(0.5,-0.35)},
            anchor=north,
            legend columns=-1,
            font=\footnotesize,
            /tikz/every even column/.append style={column sep=0.5cm}
        },
        restrict y to domain*=0:225, 
    visualization depends on=rawy\as\rawy, 
    after end axis/.code={ 
            \draw [ultra thick, white, decoration={snake, amplitude=2pt}, decorate] (rel axis cs:0,1.01) -- (rel axis cs:1,1.01);
        },
        nodes near coords={%
            \pgfmathprintnumber{\rawy}
        },
        y label style = {xshift=-0.2em,font=\footnotesize},
         yticklabel style = {xshift=-0.2em,font=\footnotesize},
        symbolic x coords={
            RPi3B,RPi4, Jetson, EGS, ES Tiny, ES Medium, ES Large, AWS t2.micro, AWS c5.large, AWS m5a.xlarge},
       xticklabel style = {rotate=45, xshift=-0.2em,font=\footnotesize},
axis lines*=left,
         clip=false,
         area legend
    ]
    \addplot [color =blue, fill=blue!30] coordinates {
      (RPi3B, 328.1 ) 
      (RPi4, 801.1 )
      (Jetson, 448.3)
      (EGS, 813.3)
      (ES Tiny, 55.1) 
      (ES Medium, 65.3)
      (ES Large,61.2) 
      (AWS t2.micro, 26.9)
      (AWS c5.large, 24.8) 
       (AWS m5a.xlarge, 25.2)
       };

    \addplot [color =red, fill=red!30] coordinates {
      
      (RPi3B, 0.9) 
      (RPi4, 0.98 )
      (Jetson, 1.2)
      (EGS,0.94)
      (ES Tiny, 7.1) 
      (ES Medium, 7.0)
      (ES Large, 7.0) 
      (AWS t2.micro, 22.1)
      (AWS c5.large, 20.7) 
       (AWS m5a.xlarge, 20.1)
       };
 
    \legend{Effective throughput (\SI{}{\mega\bit\per\second}), Latency (\SI{}{\milli\second})}
        \end{axis} 
  \end{tikzpicture}
    \caption{Round-trip communication latency and effective throughput measured with iPerf3 and ICMP echo request.}
    \label{fig:network}
\end{figure}
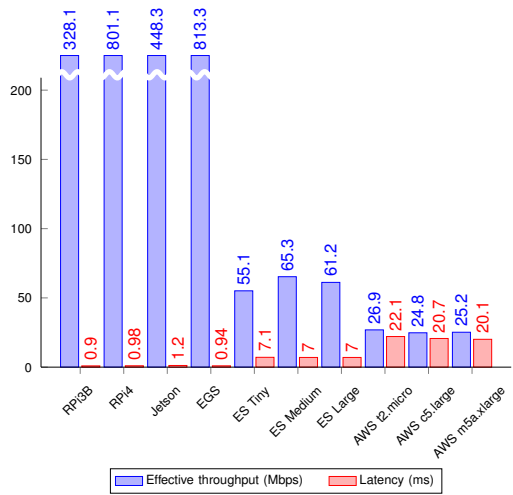

Figure~\ref{fig:network} shows the average results with a standard deviation between $0.5\%$ for EGS to $15\%$ for the Exoscale \texttt{Tiny} instance. The single-board computers and edge devices provide a $10$ times higher throughput and $20$ times lower latency.

\subsubsection{Recommendation.} \highlight{The edge and fog resources are most suitable for applications that generate frequent input and output requests with larger data sizes.}

\subsection{4.5 Carbon emission}
We evaluate the power consumption of the physical devices used for the convolutional neural network training in TensorFlow.
We use a digital multimeter to physically measure the average electrical current during training on the edge and fog resources. We rely on an AWS research report to approximate the power consumption of the fog devices and cloud instances provided by AWS and Exoscale for different utilization rates~\cite{a4}. We estimate the carbon emission directly correlated with the power consumption~\cite{AA1}, based on the grams of $CO_2$ emissions for producing one \SI{}{\kilo\watt\hour} of energy in the European Union.

\begin{figure}[h!]
    \centering
\begin{tikzpicture}[scale=0.6]
  \centering
  \begin{axis}[
        ybar, axis on top,
        height=8cm, width=12cm,
        bar width=0.4cm,
        nodes near coords,
        every node near coord/.append style={rotate=90, anchor=west},
       major grid style={draw=black},
        enlarge y limits={value=.1,upper},
        ymin=0, ymax=6,
        legend style={
            at={(0.5,-0.35)},
            anchor=north,
            legend columns=-1,
            font=\footnotesize,
            /tikz/every even column/.append style={column sep=0.5cm}
        },
        restrict y to domain*=0:117, 
    visualization depends on=rawy\as\rawy, 
    after end axis/.code={ 
            \draw [ultra thick, white, decoration={snake, amplitude=2pt}, decorate] (rel axis cs:0,1.01) -- (rel axis cs:1,1.01);
        },
        nodes near coords={%
            \pgfmathprintnumber{\rawy}
        },
        ylabel={Carbon emission [\SI{}{\gram}]},
        y label style = {xshift=-0.2em,font=\footnotesize},
         yticklabel style = {xshift=-0.2em,font=\footnotesize},
        symbolic x coords={
            RPi3B, RPi4, Jetson, EGS, ES Tiny, ES Medium, ES Large, AWS t2.micro, AWS c5.large, AWS m5a.xlarge},
       xticklabel style = {rotate=45, xshift=-0.2em,font=\footnotesize},
       axis lines*=left,
         clip=false,
         area legend
    ]
    \addplot [color =blue, fill=blue!30] coordinates {
     
      (RPi3B, 1.0)
       (RPi4, 0.7)
       (Jetson, 0.71)
      (EGS, 1.5)
      (ES Tiny, 2.3) 
      (ES Medium, 4.1)
      (ES Large, 3.9) 
      (AWS t2.micro, 2.4)
      (AWS c5.large,3.8) 
       (AWS m5a.xlarge, 5.4)
       };
    \legend{Carbon emissions}
  \end{axis}
  
  \end{tikzpicture}
    \caption{Carbon footprint for training a neural network with accuracy of above $80\%$.}
    \label{fig:carbon}
\end{figure}
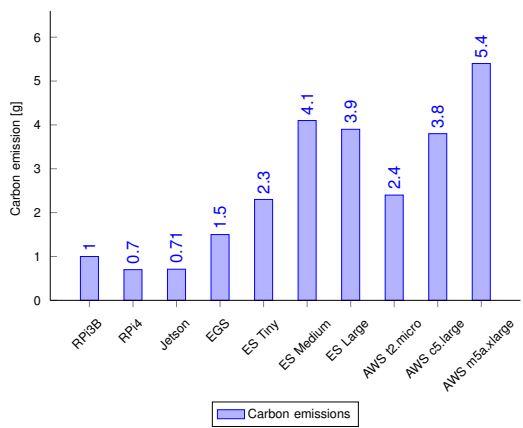

Figure~\ref{fig:carbon} shows that the edge devices  emit up to six times less carbon during training. We therefore expect to reduce the carbon emissions by \SI{1000}{\kilo\gram} per year by offloading the computation from the cloud to the edge, which is equivalent to a \SI{5517}{\kilo\meter}-long travel with a gasoline vehicle. 

\subsubsection{Recommendation.} We recommend offloading applications with soft execution time constraints (e.g., video-on-demand, data filtering, model training with small data-sets) to the edge devices. This reduces the energy costs of service providers and the carbon footprint.

\section{5 Conclusion}
\highlight{In this article we provide a set of recommendations for practitioners on where to offload their applications across the computing continuum, summarized in Table \ref{tbl:recomend}. 
We formulate the recommendations based on a systematic performance and carbon footprint analysis of a selected set of applications on a heterogeneous set of devices and cloud instances across the computing continuum. For this purpose, we deployed a representative testbed called Carinthian Computing Continuum that spawns across a three-layered conceptual architecture. Our results revealed that to reduce the network traffic over the computing continuum it is recommended to offload to edge and fog resources, while we advocate the cloud for lower execution times. Lastly, for decreasing the $CO_2$ emissions, with an acceptable computational performance penalty, we recommend edge resources.
}

\begin{table*}[t]
\scriptsize
\caption{\highlight{Recommendations for application offloading across the computing continuum.}}
\centering
\color{dgreen}
\begin{tabular}{|c|c|c|c|}
\hline
\backslashbox[23mm]{\textit{Application}}{\textit{Requirement}} &
 \textit{Low network load} & \textit{Low execution time}& \textit{Low $CO_2$ emissions}  \\
\hline
\textit{Video encoding} & Edge/Fog & Cloud  & Edge  \\
\hline
\textit{Machine learning} & Edge & Cloud/Fog & Edge  \\
\hline
\textit{In-memory analytic} & Cloud/Fog & Cloud & Edge    \\
\hline
\end{tabular}
\label{tbl:recomend}
\end{table*}

\begin{IEEEbiography}{Dragi Kimovski}{\,}is a tenure-track researcher at the Institute of Information Technology (ITEC), Klagenfurt University. He earned his PhD in 2013 from Technical University of Sofia (Bulgaria). He was assistant professor at the University for Information Science and Technology in Ohrid (North Macedonia) and senior researcher at the University of Innsbruck (Austria). His research interests include fog and edge computing, multi-objective optimization, and distributed storage. 
\end{IEEEbiography}

\begin{IEEEbiography}{Roland Math\'a}{\,}is a PhD student with an
MSc degree in computer science from the University of Innsbruck (Austria) in 2014. His interests include cloud simulation, workflows, and multi-objective optimization.
\end{IEEEbiography}

\begin{IEEEbiography}{Josef Hammer}{\,}studied computer science in Austria and Australia and received his MSc degree from the Klagenfurt University. After having spent ten years at CERN and in the automotive industry, he currently pursues a PhD degree in computer science at ITEC, Klagenfurt University. His research focus is on edge computing in connection with 5G mobile networks.
\end{IEEEbiography}

\begin{IEEEbiography}{Narges Mehran}{\,}is a PhD student at ITEC, Klagenfurt University. She received her MSc degree in computer architecture from the University of Isfahan (Iran) in 2016. Her research interests include cloud, fog and edge computing for future Internet applications.
\end{IEEEbiography}

\begin{IEEEbiography}{Hermann Hellwagner}{\,}is professor at and the chair of ITEC, Klagenfurt University. He received his PhD degree from the University of Linz (Austria) in 1988. His research interests include distributed multimedia systems, information-centric networking, edge computing, and communication in UAV swarms.
\end{IEEEbiography}

\begin{IEEEbiography}{Radu Prodan}{\,}is professor in distributed systems at ITEC, Klagenfurt University. He received his PhD degree in 2004 from the Vienna University of Technology and was Associate Professor until 2018 at the University of Innsbruck (Austria). His research interests include performance and resource management tools for parallel and distributed systems.
\end{IEEEbiography}

\end{document}